\documentclass[aps,twocolumn,pra,superscriptaddress,notitlepage,floatfix]{revtex4}
\usepackage{amssymb, graphicx,subfigure}
\usepackage{enumerate}
\usepackage{amsmath,bm}
\usepackage{color}
\usepackage{natbib} 
\usepackage[latin1]{inputenc}
\usepackage{hyperref}

\newcommand{\degree}{\ensuremath{{}^\circ}}

\newcommand{\etal}{et al}
\newcommand{\ie}{i.\,e.}

\renewcommand{\figurename}{\footnotesize Fig.}

\begin{document}
\makeatletter
\def\subsubsection{\@startsection{subsubsection}{3}{10pt}{-1.25ex plus -1ex minus -.1ex}{0ex plus 0ex}{\normalsize\bf}}
\def\paragraph{\@startsection{paragraph}{4}{10pt}{-1.25ex plus -1ex minus -.1ex}{0ex plus 0ex}{\normalsize\textit}}
\renewcommand\@biblabel[1]{#1}
\renewcommand\@makefntext[1]%
{\noindent\makebox[0pt][r]{\@thefnmark\,}#1}
\DeclareRobustCommand\onlinecite{\@onlinecite}
\def\@onlinecite#1{\begingroup\let\@cite\NAT@citenum\citealp{#1}\endgroup}
\def\equationautorefname{equation}%
\def\tagform@#1{\maketag@@@{\ignorespaces#1\unskip\@@italiccorr}}
\let\orgtheequation\theequation
\def\theequation{(\orgtheequation)}
\makeatother
\renewcommand{\figurename}{\small{Fig.}~}

\title{Theoretical description of adiabatic laser alignment and mixed-field
         orientation: the need for a non-adiabatic model}

\author{J.\ J.\ Omiste}
\affiliation{Instituto Carlos I de F\'{\i}sica Te\'orica y Computacional,
and Departamento de F\'{\i}sica At\'omica, Molecular y Nuclear,
  Universidad de Granada, 10871 Spain} 

\author{M.\ G\"arttner}
\affiliation{Max Planck Institut f\"ur Kernphysik, Saupfercheckweg 1, 69117
      Heidelberg, Germany.}
       
\author{P.\ Schmelcher}
\affiliation{Zentrum f\"ur Optische Quantentechnologien, Universit\"at Hamburg,
      Luruper Chaussee 149, Hamburg, 22761, Germany.}

\author{R.\ Gonz\'{a}lez-F\'{e}rez}
\affiliation{Instituto Carlos I de F\'{\i}sica Te\'orica y Computacional,
and Departamento de F\'{\i}sica At\'omica, Molecular y Nuclear,
  Universidad de Granada, 10871 Spain}

\author{L.\ Holmegaard}
\affiliation{Department of Chemistry, University of Aarhus, 8000 Aarhus C,
      Denmark.}

\affiliation{Center for Free-Electron Laser Science, DESY, 
Notkestrasse 85, 22607  Hamburg, Germany}

\author{J.\ H.\ Nielsen}
\affiliation{
Department of Physics and Astronomy, University of Aarhus, 8000 Aarhus C, Denmark.}

\author{H.\ Stapelfeldt}
\affiliation{Department of Chemistry, University of Aarhus, 8000 Aarhus C,
      Denmark.}
\affiliation{Interdisciplinary Nanoscience Center (iNANO), University of Aarhus,
      8000 Aarhus C, Denmark.}

\author{J.\ K\"upper}
\affiliation{Center for Free-Electron Laser Science, DESY, Notkestrasse 85, 22607
      Hamburg, Germany}
\affiliation{Fritz-Haber-Institut der MPG, Faradayweg 4--6, 14195 Berlin, Germany}
\affiliation{University of Hamburg, Luruper Chaussee 149, 22761 Hamburg, Germany}

\date{\today}

\begin{abstract}
      We present a theoretical study of recent laser-alignment and mixed-field-orientation
      experiments of asymmetric top molecules. In these experiments, pendular states were created
      using linearly polarized strong ac electric fields from pulsed lasers in combination with weak
      electrostatic fields. We compare the outcome of our calculations with experimental results
      obtained for the prototypical large molecule benzonitrile (C$_7$H$_5$N) [J.\,L.\ Hansen \etal,
      \textit{Phys.\ Rev.} A, \textbf{83}, 023406 (2011)] and explore the directional properties of
      the molecular ensemble for several field configurations, \ie, for various field strengths and
      angles between ac and dc fields. For perpendicular fields one obtains pure alignment, which is
      well reproduced by the simulations. For tilted fields, we show that a fully adiabatic
      description of the process does not reproduce the experimentally observed orientation, and it
      is mandatory to use a diabatic model for population transfer between rotational states. We
      develop such a model and compare its outcome to the experimental data confirming the
      importance of non-adiabatic processes in the field-dressed molecular dynamics.
\end{abstract}

\maketitle

\section{Introduction}
\label{sec:intr}


Controlling molecular motions has direct impact in a wide variety of molecular sciences, including
stereo-chemistry~\cite{brooks:jcp45, Stolte:BBGPC86:413, zare:science, aquilanti:pccp_7},
molecular-frame investigations of geometric and electronic properties, such as photoelectron angular
distributions~\cite{Landers:PRL87:013002, Holmegaard:natphys6,bisgaard:science323} and high-harmonic
generation~\cite{PhysRevLett.99.243001,velotta:phys_rev_lett_87}, as well as for diffractive imaging
of gas-phase molecules~\cite{PhysRevLett.92.198102,Filsinger:PCCP13:2076}, aiming at recording the
``molecular movie''~\cite{Miller:ActaCrystA66:137}. Recently, there has been tremendous progress in
the control of the translational~\cite{Meerakker:NatPhys4:595} and
rotational~\cite{stapelfeldt:rev_mod_phys_75_543, kupper:prl102, Ghafur:NatPhys5:289,rouzee:new_j_phys_11} motions of
even complex molecules. For extremely well controlled ultracold alkali dimers, direct quantum
effects in the stereodynamics of molecular reactions have recently been
observed~\cite{Miranda:NatPhys}.

Angular confinement of molecular ensembles is referred to as alignment -- the confinement of
molecule-fixed axes along laboratory-fixed axes -- and orientation -- adding a well-defined
direction. Traditionally, these two levels of angular control have been separated: strong ac fields
from pulsed lasers have been used to create alignment~\cite{stapelfeldt:rev_mod_phys_75_543},
whereas state-selection~\cite{Reuss:StateSelection} and brute-force orientation using strong dc
electric fields~\cite{loesch:jcp93, friedrich:zpd18} have been used to create orientation (which typically also creates alignment). About a decade ago it was realized that strong
simultaneous alignment and orientation could be created using combined ac and dc electric
fields~\cite{friedrich:jcp111,friedrich:jpca103}. This has been experimentally verified in a few
cases~\cite{baumfalk:jcp114, sakai:prl_90, PhysRevA.72.063401}. Adding methods to control the
translational motion one can perform a quantum-state selection before the alignment and orientation
experiment~\cite{Meerakker:NatPhys4:595, Schnell:ACIE48:6010, Kuepper:FD142:155}. This two-step
approach has allowed the creation of unprecedented degrees of one-dimensional (1D) and
three-dimensional (3D) alignment and orientation even for complex asymmetric top
molecules~\cite{kupper:prl102, kupper:jcp131, Nevo2009, Holmegaard:natphys6, PhysRevA.83.023406}.

Theoretical studies of the rotational spectra in the presence of combined electrostatic and
non-resonant radiative fields have been restricted to linear and symmetric top molecules so
far~\cite{friedrich:jpca103, friedrich:jcp111, hartelt_jcp128}. Recently, some of us developed the
theory for asymmetric top molecules in combined fields~\cite{omiste:2011}. When the static and
linearly polarized laser field directions are tilted with respect to one another, the symmetries of
the corresponding Hamiltonian are significantly reduced. This is the most general field
configuration, its theoretical treatment being most challenging. Since each external field
interaction breaks different symmetries of the field-free Hamiltonian, the order in which the fields
are turned on determines the evolution of the field-dressed rotational dynamics. The labels of the
eigenstates at a certain field configuration obtained by adiabatic following depend on the path
through parameter space, \ie, monodromy is observed. For linear and symmetric top molecules exposed
to external fields, the phenomena of classical and quantum monodromy have been encountered in the
corresponding dressed spectra~\cite{arango:ijbc1127, arango:j_chem_phys_122, kozin:jcp118}. The
labeling procedure is also numerically very demanding due to the large amount of genuine and avoided
crossings occurring between adjacent states as one of the field parameters is varied.

In the present work, we describe the recent experimental results obtained for the alignment and
orientation of asymmetric top molecules in combined fields~\cite{kupper:prl102, kupper:jcp131,
   Nevo2009, Holmegaard:natphys6, PhysRevA.83.023406}. In the framework of the rigid rotor
approximation, we perform a full investigation of the field-dressed eigenstates for those field
configurations considered in the experiment. The Schr\"odinger equation for the rotational/pendular
states is solved separately for each irreducible representation by expanding the wave function in a
basis with the correct symmetries. Our theoretical analysis also includes i) a diabatic approximation to
account for population transfer through the avoided crossings as the laser intensity is varied; ii)
the velocity distribution of the ions after the Coulomb explosion; and iii) a volume effect model to
describe the fact that not all the molecules feel the same laser intensity because the orienting and
the detection laser pulses have finite spatial intensity profiles. The field-dressed eigenfunctions
are weighted with the known relative state populations in the molecular beam~\cite{kupper:jcp131}.
Then, we compute the angular probability density functions for different field configurations and
their velocity-mapping images (VMIs). The recently performed alignment and orientation experiment
for benzonitrile (BN, C$_7$H$_5$N) molecules~\cite{Holmegaard:natphys6, PhysRevA.83.023406} provides us
with experimental data that is very well suited to present and discuss our theoretical model in a
comparative study. We numerically compute the alignment and orientation for an ensemble of
quantum-state selected benzonitrile molecules, and compare our results to the experimental data.
For comparison with a cold thermal ensemble -- without state selection -- we also
   provide the corresponding results for a benzonitrile sample at 1~K. In particular, for
perpendicular fields, we obtain good agreement between the computational and experimental results
for the degree of alignment. For tilted fields, we show that a fully adiabatic description
of the rotational/pendular dynamics cannot reproduce the experimental results for the 
mixed-field orientation of benzonitrile. When a diabatic model is implemented for the treatment of
the avoided crossings, our theoretical study reproduces with reasonable accuracy the experimental
degree of orientation. Hence, we demonstrate the impact of non-adiabatic processes on the
field-dressed molecular dynamics. We have developed a general theoretical description of alignment
and mixed-field orientation for asymmetric tops in long pulses of strong ac electric and weak dc
fields.

The paper is organized as follows: In Sec. \ref{sec:exp} the relevant experimental details are
described. The theoretical model is presented in Sec. \ref{sec:theo}, which includes the discussion
of the rigid rotor Hamiltonian and its symmetries, the diabatic model to treat the avoided
crossings, the screen projection of the 3D probability densities and the experimental observables.
The theoretical results for the alignment and orientation of a beam of benzonitrile molecules are
compared to the experimental data in Sec. \ref{sec:comp}. The conclusions and outlook are provided in
Sec. \ref{sec:dis}.

\section{Experimental details}
\label{sec:exp}

A detailed description of the alignment and mixed-field orientation experiments is given
elsewhere~\cite{kupper:jcp131, Holmegaard:natphys6, PhysRevA.83.023406}. Briefly, a
pulsed, cold molecular beam of benzonitrile molecules seeded in helium is expanded from an
Even-Lavie valve into vacuum. The molecular beam is skimmed before entering a $15$~cm long
electrostatic deflector $41$~cm downstream from the nozzle. The deflector disperses the molecules in
the beam according to their effective dipole moments, creating a vertically varying distribution of
quantum states in the probe region $77$~cm downstream from the nozzle. Alignment and orientation is
induced by the dc electric field of the VMI spectrometer and by a strong Nd:YAG laser pulse ($10$~ns,
$1064$~nm) and probed using ion imaging of CN$^+$ fragments following Coulomb explosion of
benzonitrile 
with a strong ultrashort Ti:Sapphire laser pulse ($30$~fs, $800$~nm,
$5.4\times10^{14}$~W/cm$^2$). Alignment and orientation experiments were performed for the
undeflected beam and for the quantum-state selected sample at a vertical height of $1.75$~mm; see
Figure~2 of reference~\onlinecite{PhysRevA.83.023406} for details.

\section{Theoretical model}
\label{sec:theo}

An exact theoretical quantum description of the alignment and orientation experiments is very
demanding, since it requires the solution of the corresponding time-dependent Schr\"odinger equation
for each state populated in the molecular ensemble. Instead, we retreat to a quasi-static
description parametric in the field strength and angles, and solve the time-independent
Schr\"odinger equation for several field configurations. A diabatic model, based on the field-free
symmetries, is used to account for the population transfer through the avoided crossings encountered
as the YAG pulse intensity is varied. 

\subsection{The Hamiltonian}
\label{subsec:ham}

We consider a nonresonant laser field of intensity $I$ and linearly polarized along the $Z_L$-axis
of the laboratory fixed frame (LFF), $(X_L,Y_L,Z_L)$, and an homogeneous electrostatic field of
strength $E_S$ contained in the $X_LZ_L$-plane and forming an angle $\beta$ with $Z_L$. The rigid
rotor Hamiltonian of a polar asymmetric top molecule exposed to this field
configuration is given by
\begin{eqnarray}
   \nonumber
   H &=& J_{X_M}^2B_{X_{M}}+J_{Y_M}^2B_{Y_{M}}+ J_{Z_M}^2 B_{Z_{M}}
-E_S\mu \cos\theta_S\\
   &&-\cfrac{2\pi I}{c} (\alpha^{Z_MX_M}\cos^2\theta+\alpha^{Y_MX_M}\sin^2\theta\sin^2\chi),
\label{eqn:hamiltonian}
\end{eqnarray}
with $B_{X_{M}}$, $B_{Y_{M}}$, and $B_{Z_{M}}$ being the rotational constants. The molecule or body
fixed frame (MFF) $(X_{M},Y_{M},Z_{M})$ is defined so that the permanent electric dipole moment
$\mu$ is parallel to the $Z_{M}$-axis. The LFF and the MFF are related by the Euler angles
$(\phi,\;\theta,\;\chi)$\cite{zare}. The polarizability tensor is diagonal in the MFF with
components $\alpha_{ii}$ with $i=X_{M},\;Y_{M},\;Z_{M}$, and the interaction with the laser field
depends on the polarizability anisotropies $\alpha^{ji}=\alpha_{jj}-\alpha_{ii}$,
$i,j=X_{M},Y_{M},Z_{M}$. The angle between the static electric field and the molecular $Z_{M}$-axis
is $\theta_S$ with  $\cos\theta_S=\cos\beta\cos\theta+\sin\beta\sin\theta\cos\phi$.
This study is restricted to molecules with the permanent dipole moment parallel to one of the
principal axis of inertia, as it is the case for benzonitrile.

Let us shortly summarize the approximations and assumptions made in the derivation of the
Hamiltonian~\eqref{eqn:hamiltonian}. We perform a non-relativistic description within the framework
of the rigid rotor approximation, assuming that the electronic and vibrational structures are not
affected by the external fields. In addition, we presuppose that the laser is non-resonant and that
the inverse of the oscillation frequency is much larger than the rotational period of the molecular
system and pulse duration. Thus, we can
average over the rapid oscillations, so that the interaction of this field with the molecular dipole
moment is zero, and only the interaction with the polarizability is left. Additionally, since the
YAG laser pulse duration is much larger than the time scale of the rotational dynamics, we assume
that the alignment and orientation processes are, in principle, adiabatic, and we take the time
profile of the pulse as a constant equal to $1$. The validity of this assumption will be discussed
in detail through the paper. The spatial dependence of the laser intensity will be also taken into
account by means of a volume effect model, see below.

The configuration of the fields determines the symmetries of the
Hamiltonian~\eqref{eqn:hamiltonian}. For the field-free case, they are given by the spatial group
SO(3) and the molecular point group D$_2$ -- \ie, the Fourgroup V -- consisting of the identity and
the two-fold rotations $C_2^{i}$ around the MFF $i$-axis, with $i=X_M,\,Y_M$ and $Z_M$. The
Schr\"odinger equation associated with the Hamiltonian~\eqref{eqn:hamiltonian} can not be solved
analytically, and only the total angular momentum, $J$, and its projection onto the $Z_L$-axis of
the laboratory fixed frame, $M$, are good quantum number, whereas $K$, the projection of
$\mathbf{J}$ on the $Z_M$-axis of the molecular fixed frame, is not conserved. Since an external
field defines a preferred direction in space, the symmetries of the corresponding Hamiltonian are
reduced compared to the field-free case. Here we consider three field-configurations. For a single
static field parallel to the LFF- $Z_L$-axis, the symmetry operations are $C_2^{Z_M}$, any arbitrary
rotation around the $Z_L$-axis and a reflection in any plane containing $Z_L$, thus, $M$ is still
conserved. For a certain $|M|$, there are $4$ irreducible representation, and for $M\ne0$ the states
with $M$ and $-M$ are degenerate. If the molecule is exposed to both fields, with the electric field
being rotated away from the $Z_L$-axis, the azimuthal symmetry is lost and $M$ ceases to be a good
quantum number. In the perpendicular case, $\beta=90^{\circ}$, the Hamiltonian
\eqref{eqn:hamiltonian} is invariant under $C_2^{Z_M}$, a rotation of $\pi$ around the $X_L$-axis,
and the reflection $\sigma_{X_LZ_L}$ ($X_LZ_L$ is the plane containing the fields); in consequence
there are $8$ irreducible representations. For tilted fields with $\beta \ne90^{\circ}$, two
symmetries are left, $\sigma_{X_LZ_L}$ and $C_2^{Z_M}$, and the corresponding group has only $4$
irreducible representations. For a detailed description and analysis of the symmetries for all
possible situations, we refer the reader to the recent works\cite{kanya:pra70,omiste:2011}.

The field-dressed eigenstates of an asymmetric top are characterized by the avoided crossings
appearing between levels of the same symmetry as one of the parameters of the field configuration,
i.e., $E_S$, $I$, or $\beta$, is varied. For non-collinear fields, the small amount of irreducible
representations implies an eigenstate diagram with a high degree of complexity due to the large
number of avoided crossings. Furthermore, a small degree of asymmetry on the inertia tensor
facilitates the appearance of avoided crossings on the corresponding field-dressed 
states~\cite{escribano:pra62}. For the
correct analysis of the computational results, these avoided crossings should be distinguished from
genuine ones taking place between levels of different symmetries. Hence, the time-independent
Schr\"odinger equation associated to the Hamiltonian \eqref{eqn:hamiltonian} is solved for each
irreducible representation by expanding the wave function in a basis that respects the corresponding
symmetries\cite{omiste:2011}.

The field-free states are identified by the quantum labels $J_{K_a K_c}M$, with $J$ and $M$ being
good quantum numbers, and $ K_a$ and $K_c$ the projections of $J$ onto the $Z_M$-axis of the
molecular fixed frame in the oblate and prolate limiting cases~\cite{king_jcp11}, respectively. For
reasons of addressability, we will denote the field dressed states by means of these field-free
labels indicating if an adiabatic or diabatic picture has been used.

\begin{figure}[tb]
   \centering
   \includegraphics[scale=.7]{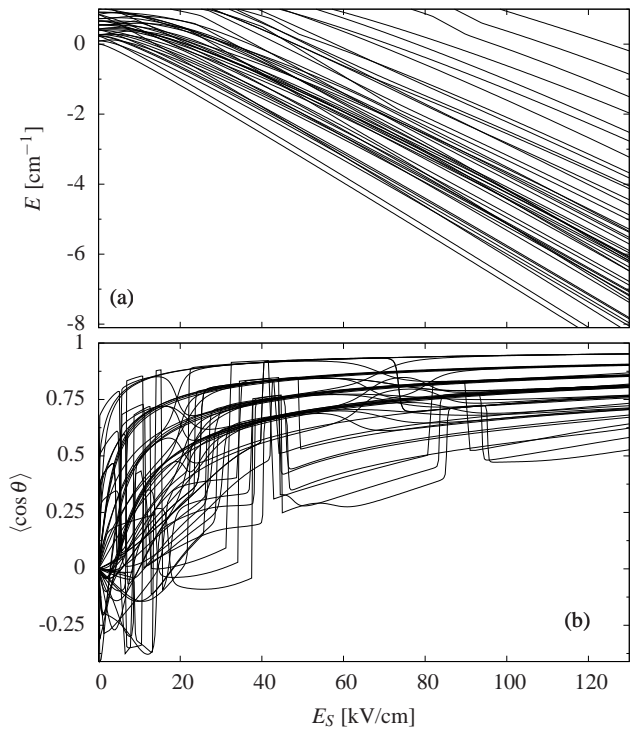}
   \caption{(a) Energies and (b) expectation value $\langle\cos\theta\rangle$ as a function of the
      static field strength of the populated rotational states for benzonitrile.}
   \label{fig:spectrum_population}
\end{figure}

To illustrate the molecular dynamics in the state-selection process using the electric deflector, we
plot the Stark energies and expectation value $\langle\cos\theta\rangle\equiv\langle J_{K_a K_c}M
|\cos\theta|J_{K_a K_c}M\rangle$ of the populated states of benzonitrile in
\autoref{fig:spectrum_population}\,(a) and (b) as a function of the electrostatic field strength.
This plot covers the range of electric field strengths present in the deflector. It includes $84$
individual rotational states accounting for $98$\,\% of the population of the molecular ensemble. This
level diagram shows a complex structure with both genuine and avoided crossings. 
In the weak electrostatic field regime,
both high- and low-field seekers are encountered, whereas in the pendular limit, and, in particular,
in the regime of interest for the state selection experiment, all these levels are
high-field-seekers. For $E_S>50$ kV cm$^{-1}$, most of these populated states present a significant
orientation with $\langle\cos\theta\rangle>0.5$, and $32\%$ of them have a strong one with
$\langle\cos\theta\rangle>0.8$. The avoided crossings have a strong impact on the character of the
involved states, and $\langle\cos\theta\rangle$ suffers large variations over tiny ranges of
electrostatic field strength, see \autoref{fig:spectrum_population}\,(b). Note that the dc field
applied in the orientation experiment, $286$ ~Vcm$^{-1}$, is very small on the scale of this figure.

\subsection{The diabatic model}

After entering the region of the extractor field, i.e., the static electric field of the VMI
spetrometer, the molecules are exposed
to a laser field with increasing intensity (YAG laser pulse). As the laser intensity varies, a
certain state may undergo several avoided crossings with levels of the same symmetry. The presence
of these avoided crossings as well as their diabatic or adiabatic nature have a strong impact on the
outcome of the experiment.

Several theoretical studies have analyzed in detail the character of such avoided crossings for
molecules exposed to an electrostatic field by using different adiabaticity criteria
\cite{bulthuis:jpca101, Kong:jpca104, schwettman:jcp123, escribano:pra62}. Their main and common
conclusion is that the assumption of a fully adiabatic dynamics is incorrect as no general statement
can be made about the character of the avoided crossings. These works suggest that an investigation
of the character of the avoided crossings
encountered 
as $I$ is varied should be mandatory for a correct description of the experimental results. To do
so, we use the following adiabatic passage criterion~\cite{bohm}
\begin{equation}
  \label{eq:adiabatic_passage}
  \eta = \cfrac{\left\langle i\left|\cfrac{\partial H'}{\partial t}\right|j\right\rangle}{(E_i-E_j)^2}
  \ll 1,
\end{equation}
where $E_i$ and $E_j$ are the eigenenergies for the states $i$ and $j$ and $H'$ is the interaction
term. Due to the large amount of avoided crossings in the laser field-dressed states, a systematic
and detailed study of the adiabatic or diabatic character of all of them is unfeasible. Thus, we
employ a simple diabatic model that provides an approximation to the dynamics, and determines the
population transfer as the field parameters are varied. Note, that some previous works have proposed
different population transfer models based on symmetry
considerations~\cite{bulthuis:jpca101,Kong:jpca104,kanya:j_chem_phys_121}.

If the fields are not collinear and the polarization of the laser pulse is parallel to the
$Z_L$-axis, the electrostatic field induces the coupling of states with different field-free values
of the quantum number $M$. For the experimentally used electrostatic field strength, $E_S=286$ V
cm$^{-1}$, the Stark interaction is much weaker than the laser field one. Thus, the hybridization of
the quantum number $M$ is so small that for a certain level $\langle M^2\rangle$ remains almost
unperturbed and equal to its field free value. Our diabatic model consists in assuming that i)
an avoided crossing  between two levels with different field-free values of $M$
is considered as being crossed diabatically; and ii)
crossings between levels with the same field-free value of $ M$ are passed adiabatically.
\begin{figure}[tb]
  \centering
  \includegraphics[scale=0.6]{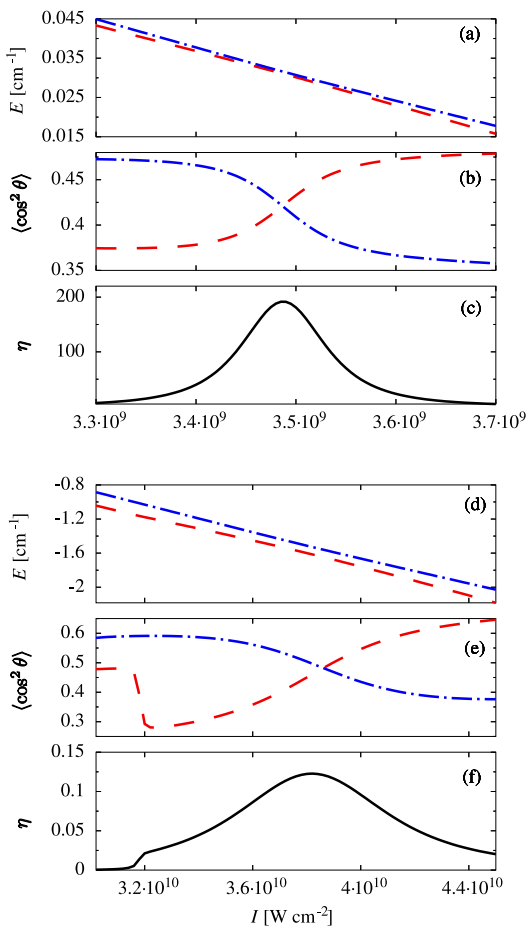}
  \caption{Energy $E$ ((a) and (d)), expectation value $\langle\cos^2\theta\rangle$ ((b) and (e)),
     adiabatic passage criterion parameter \eqref{eq:adiabatic_passage} ((c) and (f)) for the levels
     $2_{02}0$ (dash-dotted line) and $2_{02}1$ (dashed line) states; and for the states $4_{04}4$
     (dash-dotted line) and $3_{21}1$ (dashed line) states as a function of the YAG laser intensity,
     for $E_S=286$ V cm$^{-1}$ and $\beta=45^{\circ}$. See the text for more details.}
  \label{fig:adiabatic_passage}
\end{figure}
To illustrate the validity of this diabatic model we have analyzed two exemplary avoided crossings
of the field-dressed spectrum with $\beta=45^{\circ}$ and $E_S=286$ V cm$^{-1}$ by means of the
adiabatic passage parameter $\eta$ of \autoref{eq:adiabatic_passage}. In this case, the interaction
term is given by $H'=(2\pi I/c) (\alpha^{ZX}\cos^2\theta+\alpha^{YX}\sin^2\theta\sin^2\chi)$. For
the time profile of the YAG pulse, we use $I(t)=I\exp(-t^2/(2\sigma^2))$ with $I=5\times
10^{11}$~W~cm$^{-2}$ and $\sigma=4.25$~ns (FWHM=$10$~ns). \autoref{fig:adiabatic_passage}\,(a), (d),
(b), (e) and (c), (f) depict the energy, $\langle\cos^2\theta\rangle$, and $\eta$ as a function of
$I$ for the avoided crossings between the states $2_{02}0$ and $2_{02}1$, and $4_{04}4$ and
$3_{21}1$, respectively. The levels $2_{02}0$ and $2_{02}1$ suffer an avoided crossing for
$I\approx3.5\times 10^9$~W~cm$^{-2}$, with $\Delta E= 5.5\times 10^{-4}$~cm$^{-1}$ and $\eta=192$.
Hence, we conclude that it is crossed diabatically. In contrast, for the avoided crossing among the
states $4_{04}4$ and $3_{21}1$ occurring at $I\approx3.82\times 10^{10}$~W~cm$^{-2}$, we obtain
$\Delta E= 8.69 \times 10^{-2}$~cm$^{-1}$ and $\eta=0.12$, which according to
\autoref{eq:adiabatic_passage} is an intermediate case -- neither diabatic nor adiabatic. Using a
fully adiabatic picture these two states are labeled $4_{04}4$ and $3_{21}1$, but using the diabatic
model their labels are different since $\langle M^2\rangle= 3.999$ and $4.007$, respectively, for
$I=3.6\times 10^{10}$~W~cm$^{-2}$, which explains the mixing between both levels. The
criterion~\eqref{eq:adiabatic_passage} or the Landau-Zener formula do not classify this avoided
crossing as being diabatic or adiabatic, but within our approximation we consider it to be
adiabatic.

For tilted fields ($\beta\ne90\degree$), a certain avoided crossings may involve two states one
being oriented and the other one antioriented. Hence, considering it as diabatic or adiabatic has
significant consequences on the final result for orientation. Of course, our diabatic model is an
approximation, since many avoided crossings are encountered that do not fall clearly into the class
of adiabatic or diabatic crossings according to our adiabatic passage criterion. Generally, these
cases would require to solve the time-dependent Schr\"odinger equation including the time profile of
the YAG pulse.

\begin{figure}
  \centering
  \includegraphics[scale=.7]{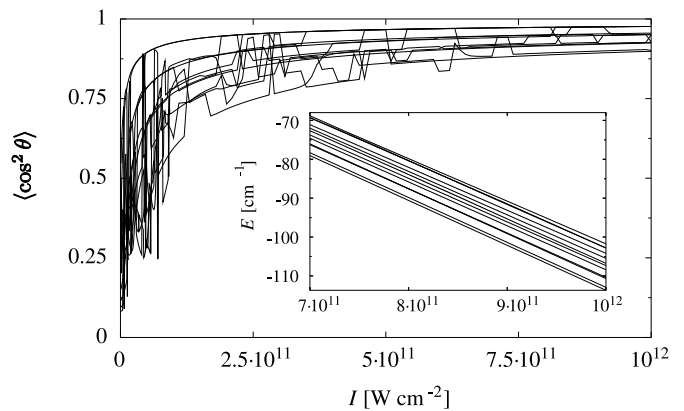}
  \caption{Expectation value $\langle\cos^2\theta\rangle$ and energies (inset panel) as a function
     of the YAG pulse intensity of the populated rotational states for benzonitrile, for $E_S=286$ V
     cm$^{-1}$ and $\beta=90^{\circ}$. Only one irreducible representation is presented.}
  \label{fig:spectrum_population_laser}
\end{figure}

If the fields are perpendicular, the corresponding Hamiltonian~\eqref{eqn:hamiltonian} has $8$
irreducible representations, and the amount of avoided crossings is significantly reduced compared
to the $\beta\ne90^{\circ}$ configuration. For the populated states of one irreducible
respresentation, the expectation value $\langle\cos^2\theta\rangle$ is plotted on
\autoref{fig:spectrum_population_laser} as a function of the laser intensity, for $E_S=286$ V
cm$^{-1}$ and $\beta=90^{\circ}$.
The inset of this figure shows the energy levels in the strong-laser-field regime. For the laser
intensities at which the experiment is performed, most of the populated levels are characterized by
a pronounced alignment. In particular, we find $\langle\cos^2\theta\rangle>0.8$ for all these states
and $I>4.1\times 10^{11}$~W~cm$^{-2}$. Since, the character -- regarding alignment-- of two states
that had previously suffered an avoided crossing is very similar, passing through it diabatically or
adiabatically is not crucial for the final alignment result.

\subsection{The projection of the probability density on the screen detector}

When a molecule is multiply ionized using an intense ultrashort laser pulse, it dissociates due to
Coulomb repulsion and the created ionic fragments are collected in a 2D space resolving detector.
Within the axial recoil approximation the momentum of the CN$^+$ fragments created in this process
is parallel to the C--CN bond direction before ionization. In order to calculate the screen image
from the molecular wave functions we define a new reference frame $(x,y,z)$, containing the detector
screen on its $yz$-plane. This frame is obtained by rotating the LFF by an angle $90^{\circ}-\beta$
about the $Y_L$-axis. Note that the electrostatic field is parallel to the $x$-axis, and the
polarization vector of the linearly polarized YAG laser lies in the $xz$-plane, forming an angle
$\beta$ with the $x$-axis.

An asymmetric top molecule in a certain state, characterized by the wavefunction
$\Psi_\gamma(\theta,\phi,\chi)$, with $\gamma=J_{K_aK_c}M$, is traveling towards a screen (velocity
in the $x$-direction). Since the direction of the molecular dipole moment is independent of the angle
$\chi$, the probability for the molecular $Z_{M}$-axis to be oriented according to $(\theta,\phi)$
is given by integrating in $\chi$ the absolute square of the wave function. The angular distribution
$\rho_\gamma(\theta,\phi)d\Omega$ provides a measure of the amount of ions ejected into the solid
angle $d\Omega=\sin \theta \, d\theta\, d\phi$. It is related to a spatial distribution on the 2D
screen by ${\rho}_\gamma(y,z)dydz= \rho_\gamma(\theta(y,z),\phi(y,z))|J|\sin (\theta(y,z)) \, dy\,
dz$, with $J$ being the Jacobian of the transformation between the coordinates $(\theta,\phi)$ and
$(y,z)$. Assuming that all the ions have the same velocity, $v$ in absolute value, this
transformation reads as
\begin{equation}
   \label{eq:trafo_screen_euler_2}
   \begin{split}
      & y = a \sin\theta \sin\phi  \\
      & z =a (\cos\theta \sin\beta + \sin\theta \cos\phi\cos\beta),
   \end{split}
\end{equation}
where $a=v t_f$, and $t_f$ is the time of flight to reach the screen. Note that with the definition
\eqref{eq:trafo_screen_euler_2} for the $y$-coordinate, any point $(y,z)$  corresponds
to two different orientations $(\theta,\phi)$ and ($\theta,180^{\circ}-\phi$). Thus, the screen
image is the sum of two projection images, each obtained for a restricted range of $\phi$.

An important ingredient, that should be taken into account to obtain realistic screen images, is the
alignment selectivity of the probe laser. We approximate the effectivity of the dissociation process
by the factor $\cos^2\alpha$, with $\alpha$ being the angle between probe laser polarization and the
C--CN bond direction~\cite{Nevo2009}. 
Note that this approximation for modeling the probe selectivity as
$\cos^2\alpha$ is in accord with the experimental observations. Thus, the 2D screen spatial
distribution is given by
\begin{equation}
\label{eq:probability_prefactor}
{\rho}_\gamma^i(y,z) = \frac{\rho_\gamma(\theta(y,z),\phi(y,z))}{a\sqrt{a^2-y^2-z^2}} A_i(y,z)
\end{equation}
where $A_i(y,z)$ is the mentioned aligment selectivity factor, and the index $i$ indicates the probe
pulse polarization. For a linearly polarized probe beam in the $x$-direction, we obtain
$A_l(y,z)=1-y^2/a^2-z^2/a^2$, and in the $z$-direction $A_p(y,z)=z/a$. For tilted fields
($\beta\ne90^{\circ}$), a circularly polarized probe in the $xz$-plane, ensures that any molecule is
ionized and detected with the same probability independently of $\beta$, and it gives 
$A_c(y,z)=1-y^2/a^2$. The apparent singularity of \autoref{eq:probability_prefactor} at
$y^2+z^2=a^2$ will disappear when we integrate over different ion recoil velocities.

\begin{figure}
  \centering
  \includegraphics[scale=.5]{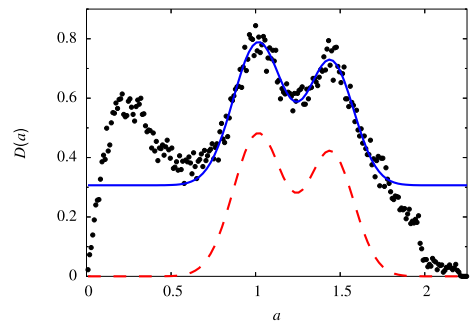}
  \caption{Experimental recoil velocity distribution of CN$^+$ ions rescaled to have a maximum for $a=1$ (points),
     fitted functions using two Gaussians (solid line) and two Gaussians without background (dashed
     line).}
  \label{fig:fit}
\end{figure}

So far we have assumed, that all ions acquire the same recoil velocity. Experimentally, however,
they follow a certain distribution $D(a)$, which is assumed to have only nonzero values at positive
velocities. The screen image is obtained by averaging over all these velocities with their
corresponding weights as
\begin{equation*}
 P_\gamma^i(y,z) = \int_0^\infty {\rho}_\gamma^i(y,z;a) D(a)\,da.
\end{equation*}
with $i=l,p$ and $c$ depending on the probe polarization. The experimental velocity distribution
$D(a)$ for the CN$^+$ ions created in the Coulomb explosion of benzonitrile is shown in
\autoref{fig:fit}. 
The second and third peak are the two  Coulomb explosion channels relevant for determining the
orientation of the C-CN axis. The two peaks are fitted to a combination of two Gaussian functions,
after subtraction of background, and used further on in the model. 
Since $a$ is not the velocity but a distance proportional to it, we have
rescaled the abscissa such that the maximum of $D(a)$ is at $a=1$.

In the experiment, the molecular beam contains molecules in different quantum states. A certain
level $\gamma =J_{K_aK_c}M$ has a relative weight $W_\gamma =W_{J_{K_aK_c}M}$ within this ensemble,
which is determined by classical trajectory simulations for all relevant rotational states using
Monte Carlo sampling of the initial phase distributions~\cite{kupper:jcp131}. Including the
resulting populations in the laser interaction zone, the final 2D-screen probability distribution
reads
\begin{equation}
   P_{\text{T}}^i(y,z) = \sum_\gamma W_\gamma P_\gamma^i(x,y),
   \label{eq:Total_projection}
\end{equation}
where the sum runs over all populated states and the index $i$ indicates the polarization of the probe.

\subsection{The experimental observables}

The alignment of a molecule is quantified by the expectation value
$\langle\cos^2\theta\rangle_\gamma$ with $0<\langle\cos^2\theta\rangle<1$, where larger values correspond
to stronger alignment. While
$\langle\langle\cos^2\theta\rangle\rangle=\sum_\gamma{}W_\gamma\langle\cos^2\theta\rangle_\gamma$ is
not experimentally determined, the alignment is characterized through
$\langle\langle\cos^2\theta_{2D}\rangle\rangle=\sum_\gamma{}W_\gamma\langle\cos^2\theta_{2D}\rangle_\gamma$,
where $\theta_{2D}=\arctan(z/y)$ is the angle between the $z$-axis of the screen plane and the
projection of the ion recoil velocity vector onto the detector plane. Let us remark that
$\langle\cos^2\theta_{2D}\rangle_\gamma$ is computed from the final 2D probability density
\autoref{eq:Total_projection}, whereas $\langle\cos^2\theta\rangle_\gamma$ from the 3D wave
functions.

When the linear polarization of the YAG laser is not perpendicular to the static electric field
($\beta\ne90^{\circ}$), the up/down symmetry of the 2D-images is lost, and an asymmetric
distribution appears showing a certain degree of orientation. This up/down asymmetry is
experimentally quantified by the ratio $N_\text{up}/N_\text{tot}$, with $N_\text{up}$ being the
amount of ions in the upper part of the screen plane, and $N_\text{tot}$ the total number of
detected ions. In our description, they are given by
\begin{equation}
\label{eq:Nups}
N_{up}=\int_{-\infty}^\infty\int_{z\ge0} P_{\text{T}}(y,z) dydz,
\end{equation}
and
\begin{equation*}
\label{eq:Ntotal}
N_{tot}=\int_{-\infty}^\infty\int_{-\infty}^\infty P_{\text{T}}(y,z) \,dydz.
\end{equation*}
Note that due to the normalization we have that $N_{tot}=1$.

Finally, to compare our numerical data with the experimental results, we take into account the
finite spatial width of the YAG and probe pulses. Both laser beams have a Gaussian profile with
widths of $\omega_Y=36\,\mu$m and $\omega_P=21\,\mu$m, respectively, and are overlapped in time and
space. The YAG pulse ensures the alignment or orientation of the molecules, whereas the probe pulse
is needed for detection purposes. The degree of alignment and the dissociation probability varies
with the position of a molecule in the interaction volume. Thus, we integrate over the overlap
region of the probe laser with the molecular beam considering the YAG-laser intensity. Hereby, we
have assumed a linear behavior of the dissociation efficiency on the probe laser intensity. This is
an approximation, as recent works about non resonant strong field dissociation for other systems 
have proved a $I^3$-dependence \cite{nielsen2011}. 
However, our calculations indicate that the orientation and
alignment results are not very sensitive to a variation of this dependency. We have also assumed
that the spatial profile of the molecular beam is much broader than that of the laser beams, hence,
the variations of the density of molecules can be neglected.

\section{Computational results}
\label{sec:comp}

In this section, we apply the above described approach using benzonitrile as a
prototype example. Recent experimental results for this molecule~\cite{PhysRevA.83.023406} allow us
to directly compare them to our numerical studies. The moments of inertia are given by $B_{X_{M}}=1214$
MHz, $B_{Y_{M}}=1547$ MHz, and $B_{Z_{M}}=5655$ MHz, the electric dipole moment is $\mu=4.515$~D,
and the principal moments of polarizability are $\alpha_{X_MX_M}=7.49$~\text{\AA}$^3$,
$\alpha_{Y_MY_M}=13.01$~\text{\AA}$^3$, and $\alpha_{Z_MZ_M}=18.64$~\text{\AA}$^3$
\cite{Wohlfart:JMolSpec247:119, PhysRevA.83.023406}.

\subsection{Alignment results}
\label{sub:alignment}

The experimental fields geometry consists of a weak static electric field perpendicular to the
screen with strength $E_s=286$~V~cm$^{-1}$ and a strong alignment laser linearly polarized along the
$Z_L$-axis ($z$-axis). For the probe pulse we consider the three possible polarizations: i) linearly
polarized perpendicular to the screen; ii) linearly polarized parallel to the screen; and iii)
circularly polarized in the $xz$-plane perpendicular to the screen. The computational results shown
below include the recoil velocity distribution and the volume effect.

\begin{figure}
  \centering
  \includegraphics[scale=.3]{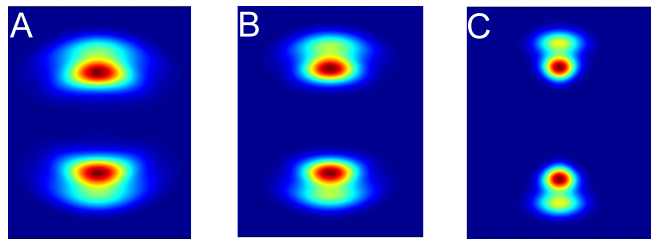}
  \caption{The 2D projection of the density distribution of the benzonitrile molecular beam for (a)
     $I=5\times 10^{10}$~W~cm$^{-2}$, (b) $10^{11}$~W~cm$^{-2}$ and (c) $7\times
     10^{11}$~W~cm$^{-2}$, $E_S=286$ V cm$^{-1}$ and $\beta=90^{\circ}$.}
  \label{fig:cos_2d_projection}
\end{figure}
The theoretical results for the density distribution of the molecular beam on the screen are
presented in \autoref{fig:cos_2d_projection}\,(a), (b) and (c) for $E_S=286$ V cm$^{-1}$ and for YAG
pulse intensities $I=5\times10^{10}$~W~cm$^{-2}$, $10^{11}$~W~cm$^{-2}$, and $7\times
10^{11}$~W~cm$^{-2}$, respectively. A strong confinement of the probability distribution is
observed, the molecules are aligned with their $Z_M$-axis pointing along the polarization axis of
the YAG laser. This 1D alignment becomes more pronounced as $I$ is increased. The two ionization
channels that characterize the velocity distribution for BN, cf.\ \autoref{fig:fit}, appear as
double maxima on the upper and lower hump of the 2D images. They become more prominent as $I$ is
increased, \ie, for a YAG pulse with $I=7\times 10^{11}$ W cm$^{-2}$, while they are overlapping for
lower intensities.

\begin{figure}
  \centering
  \includegraphics[scale=.6]{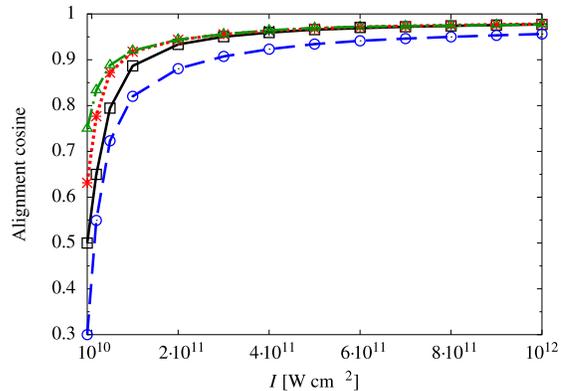}
  \caption{Alignment cosines for the molecular beam of bezonitrile as a function of the YAG pulse
     intensity, $E_S=286$ V cm$^{-1}$ and $\beta=90^{\circ}$:
     $\langle\langle\cos^2\theta\rangle\rangle$ (circles and dashed line),
     $\langle\langle\cos^2\theta_{2D}\rangle\rangle^l$ (squared and solid line),
     $\langle\langle\cos^2\theta_{2D}\rangle\rangle^c$ (asterisks and dotted line), and
     $\langle\langle\cos^2\theta_{2D}\rangle\rangle^p$ (triangles and dash-dotted line).}
  \label{fig:cos2d_bn}
\end{figure}

We quantify this alignment by the expectation values $\langle\langle\cos^2\theta\rangle\rangle$,
$\langle\langle\cos^2\theta_{2D}\rangle\rangle^l$,
$\langle\langle\cos^2\theta_{2D}\rangle\rangle^p$, and
$\langle\langle\cos^2\theta_{2D}\rangle\rangle^c$, which are presented in \autoref{fig:cos2d_bn} as
a function of the YAG intensity $I$ for the whole molecular beam.
The indexes $l$, $p$ and $c$ indicate the probe pulse polarization: linearly-polarized perpendicular
and linearly-polarized parallel to the static field and circularly-polarized, respectively. Without
the presence of any aligning or orienting fields the screen image is already biased by the
``geometric alignment'' of the probe pulse. An analytical calculation shows for a random isotropic
distribution that $\langle\langle\cos^2\theta\rangle\rangle=1/3$,
$\langle\langle\cos^2\theta_{2D}\rangle\rangle^l=0.5$,
$\langle\langle\cos^2\theta_{2D}\rangle\rangle^{p}=3/4$, and
$\langle\langle\cos^2\theta_{2D}\rangle\rangle^c=5/8$. 
This
   mild confiment obtained when the probe pulse is circularly polarized or linearly polarized
   parallel to the screen is due to the enhanced ionization probability for molecules having the
   $C_2$ axis (\ie, the C-CN axis) parallel to the laser polarization. 
If the molecules are exposed
only to the weak electric field, $E_S=286$ V cm$^{-1}$, these field-free values are only slightly
perturbed $\langle\langle\cos^2\theta\rangle\rangle\approx0.34$,
$\langle\langle\cos^2\theta_{2D}\rangle\rangle^l=0.51$,
$\langle\langle\cos^2\theta_{2D}\rangle\rangle^p=0.7496$, and
$\langle\langle\cos^2\theta_{2D}\rangle\rangle^c=0.6319$~\cite{Nevo2009}.

These four expectation values qualitatively all show a similar evolution as the
YAG pulse intensity is varied: a steep increase followed by a plateau-like behavior which indicates
a saturation of the degree of alignment. This dependence of the alignment on the YAG pulse intensity
nicely reproduces the experimental behavior observed for a molecular beam of iodobenzene (IB)
\cite{kupper:prl102, kupper:jcp131}. Note that the differences due to the polarizations of the probe
pulse are noticeable for low intensities, whereas for strong alignment fields they all converge to
the same asymptotic limit. For $I=7\times 10^{11}$ W cm$^{-2}$, we obtain
$\langle\langle\cos^2\theta\rangle\rangle=0.946$, 
$\langle\langle\cos^2\theta_{2D}\rangle\rangle^l=0.972$,
$\langle\langle\cos^2\theta_{2D}\rangle\rangle^p=0.973$, and
$\langle\langle\cos^2\theta_{2D}\rangle\rangle^c=0.974$. 
The value obtained experimentally, $0.89$, for $\langle\langle\cos\theta_{2D}\rangle\rangle^p$~\cite{PhysRevA.83.023406} is
somewhat lower which we ascribe to contaminant ions, like C$_2$H$_2^+$ (with the same mass-to-charge
ratio as CN$^+$) in the images. Ions like C$_2$H$_2^+$ do not have a strong angular confinement and
therefore reduced the apparent degree of alignment in the image. For IB, a molecule which
is expected to attain an alignment degree similar to that of BN because of a similar
polarizabilty tensor, the recoil ion used, I$^+$, is more clean since there are no contaminant ions
at the mass of 
$127$. 
As a result, the observed degree of alignment is as high as $0.97$ 
for $\langle\langle\cos\theta_{2D}\rangle\rangle^l$~\cite{kupper:jcp131} in good agreement with the theoretical predictions.

For completeness, we have also considered a thermal sample of BN at $1$~K and $I=7\times
10^{11}$~W~cm$^{-2}$, obtaining $\langle\langle\cos^2\theta_{2D}\rangle\rangle^l=0.961$, which
agrees well 
with the experimental value for IB~\cite{kumarappan:194309}.

\subsection{Orientation results}

The polarization axis of the YAG-laser is now rotated about the $Y_L$-axis, and forms an angle
$\beta$ ($\beta\ne 90^{\circ}$) with the weak electrostatic field perpendicular to the screen,
$E_s=286$ V cm$^{-1}$. The orientation ratio $N_{up}/N_{tot}$ \autoref{eq:Nups} is derived for a
circularly polarized probe pulse, including the velocity distribution and the volume effect and
using the diabatic model described above.

\begin{figure}
  \centering
  \includegraphics[scale=.4]{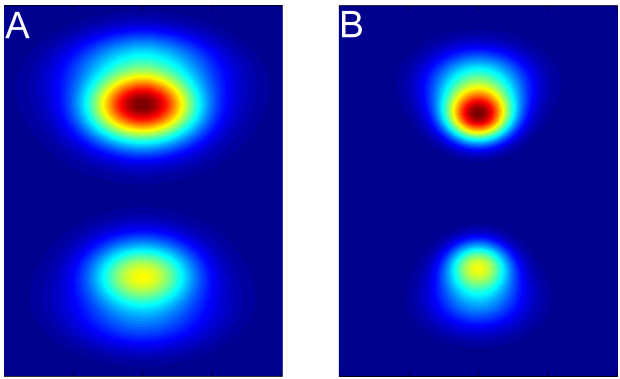}
  \caption{The 2D projection of the density distribution of the benzonitrile molecular beam for (a)
     $I=10^{11}$, and (b) $7\times 10^{11}$ W cm$^{-2}$, $E_S=286$ V cm$^{-1}$ and
     $\beta=135^{\circ}$.}
  \label{fig:orient_projection}
\end{figure}
The theoretical results for the projected density distribution of the molecular beam are presented
in Figs. \ref{fig:orient_projection}(a) and (b) for two intensities and $\beta=135^{\circ}$. As a
consequence of the rotation of the YAG polarization axis, the up/down symmetries of the wave
functions is lost. The 1D orientation becomes more pronounced as $I$ is increased. For
$\beta=135^{\circ}$, the projection of CN$^+$ ions on the 2D detector overlaps the two Coulomb
explosion channels to an extent that they cannot be discerned.

\begin{figure}[h]
  \centering
    \includegraphics[scale=.7,angle=0]{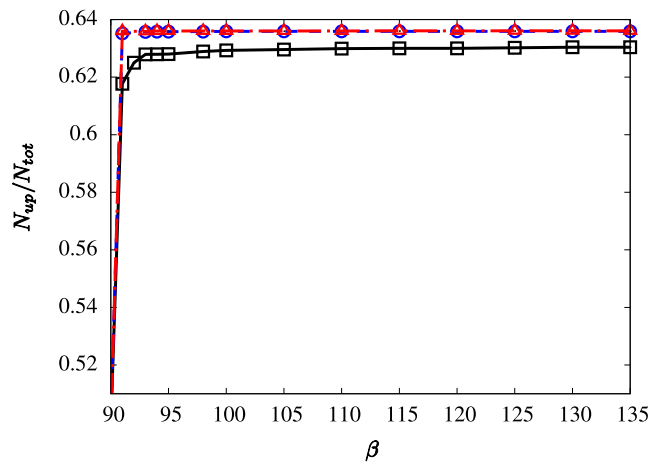}
    \caption{The theoretical orientation ratio $N_{up}/N_{tot}$ as a function of the angle $\beta$
       for the molecular beam of benzonitrile, YAG pulses with $I=10^{11}$ (squares and solid
       line), $5\times 10^{11}$ (circles and dashed line) and $7\times 10^{11}$ W cm$^{-2}$
       (triangles and dash-dotted  line) and a static field with $E_S=286$ V cm$^{-1}$.}
  \label{fig:nup_ntot_num}
\end{figure}

The theoretical results for the orientation $N_{up}/N_{tot}$ are presented in
\autoref{fig:nup_ntot_num} as a function of the tilt angle $\beta$ and for three intensities. The main feature of
the $N_{up}/N_{tot}$ is that it is almost independent of the alignment laser intensity and of
$\beta$. 
For these three intensities, all the populated states are within 
the pendular regime. Thus, if the volume effect was neglected, the ratio $N_{up}/N_{tot}$ would be
independent of $I$ for this pendular regime of laser intensities. By taking into account the spatial
distribution of the probe and YAG beams, the molecules show the smallest orientation for $I=10^{11}$
W cm$^{-2}$. This is explained by the larger contribution of lower intensities, for which the
molecules are not yet in the pendular regime, to the volume effect integral for $I=10^{11}$ W
cm$^{-2}$, and as a consequence $N_{up}/N_{tot}$ is reduced. For $I=5\times 10^{11}$ and $7\times
10^{11}$ W cm$^{-2}$, the contribution of lower intensities is not strong enough to cause any
noticeable difference.
Note that for IB, 
it was also experimentally found that  $N_{up}/N_{tot}$ is independent of the YAG pulse intensity~\cite{kupper:jcp131}.
For $\beta=90^{\circ}$, the molecular beam does not show any orientation and it holds
$N_{up}/N_{tot}=0.5$. For a certain YAG pulse intensity, $N_{up}/N_{tot}$ quickly increases as
$\beta$ is increased from $90$ and reaches a plateau-like behavior for the rest of the field
configurations. For $\beta=135^{\circ}$ and $I=$ $7\times 10^{11}$ W cm$^{-2}$, we obtain
$N_{up}/N_{tot}=0.636$ which is smaller than the experimental result of
$0.71$~\cite{PhysRevA.83.023406}. Thus, our theoretical model underestimates the orientation obtained in
the experiment.

An important discrepancy between theoretical and experimental results is the variation of
$N_{up}/N_{tot}$ for $\beta$ close to $90^{\circ}$. For a molecular beam of
IB~\cite{kupper:jcp131}, it was experimentally shown that $N_{up}/N_{tot}$ smoothly
increases (decreases) as $\beta$ is increased (decreased) from $90^{\circ}$, reaching for
$\beta\lesssim 60^{\circ}$ and $\beta\gtrsim 120^{\circ}$ the plateau. 
In
contrast, the theoretical $N_{up}/N_{tot}$ sharply increases, and its $\beta$-independent value is
already achieved for $\beta\gtrsim 91^{\circ}$. In the pendular regime, the strong laser interaction
pairs states into quasidegenerate doublets. The electrostatic field can induce a strong
coupling between these levels giving rise to a large orientation if the energy gap is small enough.
For certain intensities, some excited levels may not show such a narrow energy gap to obtain a
significant orientation, but, their relative weight within the molecular beam is so small that their
contribution to the final result is not relevant. For a linear molecule in combined
fields~\cite{friedrich:z_phys_d_1995}, such a sharp rise was predicted for the expectation value
$\langle\cos\theta\rangle$ as the static field strength is increased, i.e., increasing or decreasing
$\beta$ from $90^{\circ}$ in our case. On Figure 2 of  
Friedrich and 
Herschbach Ref. \onlinecite{friedrich:jcp111} this effect is obtained for the interaction with the laser field
being $25000$ times larger than the one with the static field. We could perform a similar comparison
and for BN, $I=7\times10^{11}$ W cm$^{-2}$, $E_S=286$ V cm$^{-1}$ and $\beta=95^{\circ}$, the
interaction of the YAG pulse, $2\pi I \alpha^{ZX}/c$, is $43600$ times larger than the coupling with
the static field, $\mu E_S \cos\beta$.

Let us remark, that our diabatic model for the population transfer is equivalent to considering a
field configuration with a linearly polarized YAG pulse being parallel to an electrostatic field
with strength $E_s\cos\beta$. This field geometry is neglecting the component of the electric field
$E_s\sin\beta$ perpendicular to the YAG polarization, which is responsible for breaking the
azimuthal symmetry and causing $M$ to stop being a good quantum number. This approximation can be
done because the interaction due to the static field is sufficiently weak. We have performed the full
calculation considering the simplified case of parallel fields, the static one with strength $E_s=
286\cos 135^{\circ}$ V cm$^{-1}$ and $I=7\times 10^{11}$ W cm$^{-2}$, obtaining for the orientation
the value $N_{up}/N_{tot}=0.636$. 

\begin{figure}
  \centering
    \includegraphics[scale=.7,angle=0]{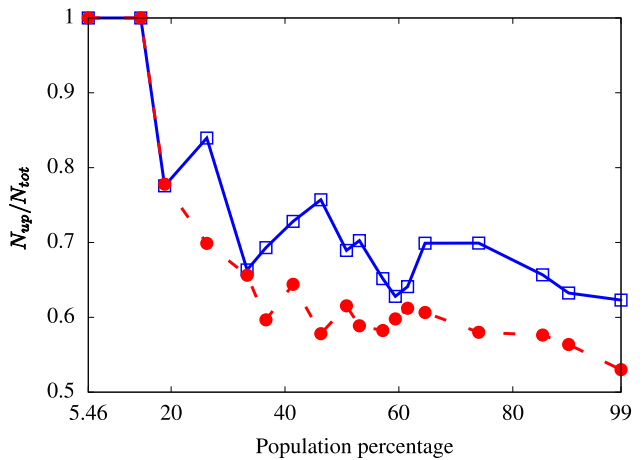}
    \caption{The theoretical orientation ratio $N_{up}/N_{tot}$ as a function of the population of
       the molecular beam of benzonitrile, for the field configuration $I=7\times
       10^{11}$~W~cm$^{-2}$, $E_S=286$~V~cm$^{-1}$, and $\beta=135^{\circ}$, computed with diabatic
       model for the population transfer (squares and solid line) and a fully adiabatic picture
       (circles and dashed line).}
  \label{fig:nup_ntot_pesos}
\end{figure}
To illustrate the necessity of the diabatic model, we build up the molecular ensemble by
successively adding states according to their weight. Thus, our initial ensemble contains only the
ground state, and for the second one, we add the second and third most populated levels. The
relative weights of the states within these molecular beams are the same as in the full molecular
ensemble. \autoref{fig:nup_ntot_pesos} presents the orientation $N_{up}/N_{tot}$ as a function of
the percentage of states included in the ensemble for $I=7\times 10^{11}$ W cm$^{-2}$ and
$\beta=135^{\circ}$. The ratio $N_{up}/N_{tot}$ has been computed with the diabatic model for the
population transfer and a fully adiabatic picture, i.e., all avoided crossings are assumed to be passed
adiabatically. The first three populated states, $0_{00}0$, $1_{01}1$ and $1_{01}-1$, are perfectly
oriented: $N_{up}/N_{tot}=1$. For the sets including the levels adding $20\%$ of the total
population, the diabatic and adiabatic results agree because these states do not suffer any avoided
crossings. As highly excited levels are added to the molecular beam, the difference between both
calculations becomes more evident due to the presence of the diabatic avoided crossings. They both
show a zig-zag decreasing trend as the population is increased. The ratio $N_{up}/N_{tot}$ computed
under the adiabaticity assumption is smaller than the diabatic result as more states are added.
Taking into account all states populated in the molecular beam, the adiabatic model does not yield
any appreciable amount of orientation: we obtain $N_{up}/N_{tot}=0.53$ which strongly underestimates
the experimental result. In addition we have observed that the adiabatic orientation ratio
$N_{up}/N_{tot}$ is no longer $\beta$-independent. This confirms that non-adiabatic crossings play a
crucial role for understanding the strong orientation observed in the experiment.

Finally, it is worth noting for the field configuration $I=7\times 10^{11}$ W cm$^{-2}$ and
$\beta=135^{\circ}$, a thermal sample of BN at $1$~K shows a weak degree of
orientation with $N_{up}/N_{tot}=0.563$.

\section{Summary and conclusions}
\label{sec:dis}

We have presented a theoretical model to investigate the degree of alignment and orientation of a
beam of asymmetric top molecules exposed to combined electrostatic and non-resonant linearly
polarized laser fields. Our analysis combines the field-dressed wave functions with the
experimental distribution of the populated quantum states. As a first step, we solve the
three-dimensional Schr\"odinger equation within the rigid rotor approximation. For a certain field
geometry, we treat each irreducible representation independently, by expanding the wave function in
a basis that respects the corresponding symmetries. Since the dc electric field strength is very
weak, we consider $M$ as being almost conserved, and a diabatic model is introduced to approximate
the population transfer trough the avoided crossings as the YAG pulse intensity is increased. The 2D
projection of a wave function is derived by using the detection selectivity factors of the probe
pulse, the velocity distribution of the detected ions, and a volume effect average.

This theoretical model has been checked by comparing the numerical and experimental results for
benzonitrile. The molecular beam has been simulated using the population of each quantum state
numerically obtained from deflection profiles~\cite{kupper:jcp131}. For several field
configurations, we have performed a detailed study of the directional properties of the molecular
mixture. In particular, we have explored the degrees of alignment and orientation as the YAG pulse
intensity and the angle between both fields are varied. For perpendicular fields, a good agreement
between the computational and experimental results is obtained. Let us remark that we do not take
into account the background of unwanted ions which contaminate the velocity mapping images for
BN and reduce the experimental degree of alignment~\cite{PhysRevA.83.023406}. Hence, a
better agreement could be achieved for other systems with cleaner Coulomb explosion imaging
channels, like IB.

Regarding the orientation results, we have shown that the assumption of a fully adiabatic dynamics
is incorrect for the prototypical experiment and predicts a non-oriented molecular beam -- which is
not in agreement with the experimental results. Indeed, we have proven that the degree of
orientation does not adiabatically follow the time envelope of the YAG laser with a FWHM$=10$ ns in
the experiment. By employing a simple diabatic model, the experimental results for orientation could
be reproduced reasonably well. Based on the comparison with the experimental measures, the important impact of
the diabaticity on the field-dressed molecular dynamics is hereby demonstrated. However, notice that
our model does not produce the smooth $\beta$-dependence of $N_{up}/N_{tot}$ that was experimentally
obtained for IB~\cite{kupper:jcp131}.

This theoretical model is based on several approximations that may be the source of discrepancies
with the experimental results. The molecules are exposed to an alignment laser pulse with a certain
time profile, but we are performing a time-independent description of the field-dressed rotational
dynamics. We compensate this deficiency by using a simple partially diabatic approach to model the
populations transfer. However, a general statement on the character of the avoided crossings can not
be made, and some of them could not be classified clearly by our adiabaticity criterion. The
semiclassical calculations of the molecular trajectories through the deflector have been derived
using a fully adiabatic picture, whereas it has been theoretically shown that such an approximation may be
   incorrect\cite{escribano:pra62}. These assumptions for both processes might not be fully satisfied. A full
time-dependent description of the process would properly treat the avoided crossing, but it is
computationally very challenging. 
Possible distortion effects due to the strong laser pulse are also not taken into account.
Furthermore, we are working within the axial recoil approximation neglecting the interactions
between the ionic fragments on their way to the detector. Analogously, collisions and interactions
between the molecules within the molecular beam~\cite{Erlekam:PCCP9:3783} were not considered. The
geometric alignment due to the strong-field ionization has been assumed to be a single-photon
absorption, but it is a multiphoton process~\cite{nielsen2011}. Nuclear hyperfine structure has been
neglected. Whereas, it can be of the same order of magnitude as the interaction with the weak static
field (\ie, for IB~\cite{Dorosh:JMolSpec246:228}), this is clearly not the case for
BN~\cite{Wohlfart:JMolSpec247:119}.

Certainly, it would be interesting to perform a comparison of these computational results with
experimental data obtained for other molecular species either with a smaller number of populated
states or with a cleaner imaging signal. A rather natural extension of the present work would be to go
beyond any of the approximations described above. A more sensitive criterion for the avoided crossing
treatment, a time-dependent description, or the inclusion of the hyperfine interaction should
improve these theoretical results.

\begin{acknowledgments}
We acknowledge support by Frank Filsinger for the trajectory simulations providing the populations 
of individual quantum states. 
We thank Gerard Meijer for continuous interest and support of this
project and Marko H\"artelt and Bretislav Friedrich for helpful discussions. Financial support by
the Spanish project FIS2008-02380 (MICINN) as well as the Grants FQM-2445 and FQM-4643 (Junta de
Andaluc\'{\i}a) is gratefully appreciated. J.J.O. acknowledges the support of ME under the program
FPU. R.G.F. and J.J.O. belong to the Andalusian research group FQM-207.

\end{acknowledgments}

   \bibliography{refs}
   \bibliographystyle{apsrev}

\end{document}